\documentclass[preprint,preprintnumbers,showpacs,aps,prd,amssymb]{revtex4-1}

\usepackage{graphicx}
\usepackage{bm}
\usepackage{amsmath}

\def\nn{\nonumber}

\def\calA{{\cal A}}

\def\calH{{\cal H}}
\def\calL{{\cal L}}
\def\calO{{\cal O}}
\def\calU{{\cal U}}

\def\Brbar{{\overline{\rm Br}}}
\def\Bbar{{\bar B}}

\def\ellbar{{\bar\ell}}
\def\hbar{{\bar h}}

\def\sbar{{\bar s}}

\def\Br{{\rm Br}}
\def\gSM{g_{\rm SM}}
\def\dU{{d_\calU}}

\begin{document}
\title{New Physics in $B_{d,s}$-$\Bbar_{d,s}$ mixings and $B_{d,s}\to\mu^+\mu^-$ decays}
\author{Jong-Phil Lee}
\email{jongphil7@gmail.com}
\affiliation{Division of Quantum Phases $\&$ Devices, School of Physics, Konkuk University, Seoul 143-701, Korea}

\begin{abstract}
A new way of probing new physics in the $B$ meson system is provided.
We define double ratios for the observables of $B_{d,s}$-$\Bbar_{d,s}$ mixings and $B_{d,s}\to\mu^+\mu^-$ decays,
and find simple relations between the observables.
By using the relations we predict the yet-to-be measured branching ratio of $B_d\to\mu^+\mu^-$ to be
$(0.809\sim1.03)\times 10^{-10}$, up to the new physics models.
\end{abstract}

\maketitle
Recent discovery of a Higgs boson at the large hadron collider (LHC) opens 
a new era of high energy physics.
It may take time to confirm whether the new particle is really the Higgs boson of the standard model (SM), 
but it looks more and more like the SM Higgs.
The discovery of the Higgs boson would mean a completion of the SM.
On the other hand, we have many reasons to believe that there must be new physics (NP) beyond the SM.
Unfortunately, the LHC up to now has not reported any clues of NP.
But it is too early to say that there is no NP at all.
$B_{d,s}$ mesons are good test beds for NP.
Especially, $B_{d,s}$-$\Bbar_{d,s}$ mixings and $B_{d,s}\to\mu^+\mu^-$ decays are loop induced phenomena in the SM 
and very sensitive to NP effects.
Current status of experiments is well compatible with the SM predictions.
For example, the LHCb and the CMS collaboration reported that \cite{LHCb1307,CMS1307}
\begin{eqnarray}
\Br(B_s\to\mu^+\mu^-)&=&(2.9^{+1.1}_{-1.0})\times 10^{-9}~,~~~
\Br(B_d\to\mu^+\mu^-)<7.4\times 10^{-10}~({\rm LHCb})~,
\label{LHCb}\\
\Br(B_s\to\mu^+\mu^-)&=&(3.0^{+1.0}_{-0.9})\times 10^{-9}~,~~~
\Br(B_d\to\mu^+\mu^-)<1.1\times 10^{-9}~({\rm CMS})~.
\label{CMS}
\end{eqnarray}
The measured value is slightly smaller than the previous LHCb measurements \cite{LHCb1211}
\begin{equation}
 \Br(B_s\to\mu^+\mu^-)=(3.2^{+1.5}_{-1.2})\times 10^{-9}~,~~~
\Br(B_d\to\mu^+\mu^-)<9.4\times 10^{-10}~.
\end{equation}
As a comparison, the SM predictions are \cite{Buras3,Isidori}
\begin{eqnarray}
\Br(B_s\to\mu^+\mu^-)&=&(3.25\pm 0.17)\times 10^{-9}~,\\
\Br(B_d\to\mu^+\mu^-)&=&(1.07\pm 0.10)\times 10^{-10}~.
\end{eqnarray}
But there is still some room for NP, as discussed in \cite{Buras3,Buras1,Buras2}.
In this paper, we provide a very simple and quick way to probe NP in 
$B_{d,s}$-$\Bbar_{d,s}$ mixings and $B_{d,s}\to\mu^+\mu^-$ decays.
The idea is that a double ratio for one observable between different flavors extracts the relevant couplings for NP, 
and they are directly related to the other observable.
Schematically, for a physical observable $\calO_i^a$ with flavor $a$,
\begin{equation}
 R_i^{ab}\equiv\frac{\calO_{i,{\rm exp}}^a/\calO_{i,{\rm SM}}^a-1}
        {\calO_{i,{\rm exp}}^b/\calO_{i,{\rm SM}}^b-1}
\simeq f_i\left(\frac{c^a}{c^b}\right)~,
\end{equation}
where $c^a$ are the new couplings and $f_i$ is some function of $c^a/c^b$.
For another observable $\calO_j$ we can also define a similar quantity $R_j^{ab}$ which would be $\simeq f_j(c^a/c^b)$.
Consequently, $R_i^{ab}$ and $R_j^{ab}$ are related through the functions $f_i$ and $f_j$,
and the relations are remarkably simplified when the new couplings belong to the category of the minimal flavor violation (MFV).
In this way, we can establish simple relations between observables of 
$B_{d,s}$-$\Bbar_{d,s}$ mixings and $B_{d,s}\to\mu^+\mu^-$ decays.
The relations are very useful because $R_i^{ab}$ and $R_j^{ab}$ are directly connected,
and the relations are different for various NP models.
For example, we can predict $\Br(B_d\to\mu^+\mu^-)$ from other known observables 
such as $\Delta M$ of $B_{d,s}$-$\Bbar_{d,s}$ mixings, 
{\em without knowing} the values of new couplings.
Or if we measure the branching ratio $\Br(B_d\to\mu^+\mu^-)$, 
we can find out from the double ratio relations which NP is realized in $B$ physics.
In this paper we specifically consider flavor changing scalar (un)particles and vector boson ($Z'$) scenarios.
Actually it was already known that $\Delta M_q$ and $\Br(B_q\to\mu^+\mu^-)$ can be related to each other \cite{Buras4}.
In our approach, $R_i^{ab}$ are directly proportional to the new physics effects,
so the resulting relations are solely those of new physics. 
The relations might be different for various models, which makes it easier to see which kind of new physics is realized.
\par
The new physics couplings adopted in this analysis are summarized as follows \cite{Buras3,jplee1}:
\begin{eqnarray}
\calL_{Z'}&=&\left[
\Delta_L^{sb}(Z')(\sbar\gamma_\mu P_Lb)+\Delta_R^{sb}(Z')(\sbar\gamma_\mu P_Rb)
+\Delta_L^{\ell\ell}(Z')(\ellbar\gamma_\mu P_L\ell)+\Delta_R^{\ell\ell}(Z')(\ellbar\gamma_\mu P_R\ell)
\right]Z'^\mu~,\label{LZ}\\
\calL_H&=&\left[
\Delta_L^{sb}(H)(\sbar P_Lb)+\Delta_R^{sb}(H)(\sbar P_Rb)
+\Delta_L^{\ell\ell}(H)(\ellbar P_L\ell)+\Delta_R^{\ell\ell}(H)(\ellbar P_R\ell)
\right]H~,\label{LH}\\
\calL_\calU&=&
\frac{c_{\calU L}^{bs}}{\Lambda_\calU^\dU}\sbar\gamma_\mu(1-\gamma_5)b~\partial^\mu\calO_\calU
  +\frac{c_{\calU L}^\ell}{\Lambda_\calU^\dU}\bar\ell\gamma_\mu(1-\gamma_5)\ell~\partial^\mu\calO_\calU~,
\label{LU}
\end{eqnarray}
where $P_{L,R}=(1\mp\gamma_5)/2$.
In $\calL_\calU$ one can also include the right-handed couplings, 
but here (and in  \cite{jplee1}) only the minimal extension of the SM are considered for simplicity.
\par
First consider the $B_{d,s}$-$\Bbar_{d,s}$ mixing.
The mixing effect is parametrized as the following quantity
\begin{equation}
 \Delta M_q=\frac{G_F^2}{6\pi^2}M_W^2m_{B_q}|V_{tb}^*V_{tq}|^2F_{B_q}^2\eta_B|S(B_q)|~,
\end{equation}
where
\begin{equation}
 S(B_q)=S_0(x_t)+\Delta S(B_q)\equiv|S(B_q)|e^{i\theta_S^{B_q}}~,
\end{equation}
and $x_t=m_t^2/m_W^2$.
Here the loop function
\begin{equation}
 S_0(x_t)=\frac{4x_t-11x_t^2+x_t^3}{4(1-x_t)^2}-\frac{3x_t^2\log x_t}{2(1-x_t)^3}~,
\end{equation}
and
\begin{equation}
 \Delta S(B_q)=[\Delta S(B_q)]_{V(S)LL}+[\Delta S(B_q)]_{V(S)RR}+[\Delta S(B_q)]_{V(S)LR}~,
\end{equation}
where the subscript $V(S)$ stands for $Z'(H)$ contributions.
Explicitly \cite{Buras1,Buras2},
\begin{eqnarray}
 [S(B_q)]_{VLL}&=&\left[\frac{\Delta_L^{bq}(Z')}{V_{tb}^* V_{tq}}\right]^2\frac{4{\tilde r}}{M_{Z'}^2\gSM^2}~,\\
 {[}S(B_q)]_{VRR}&=&\left[\frac{\Delta_R^{bq}(Z')}{V_{tb}^* V_{tq}}\right]^2\frac{4{\tilde r}}{M_{Z'}^2\gSM^2}~,\\
 {[}S(B_q)]_{VLR}&=&
   \frac{\Delta_L^{bq}(Z')\Delta_R^{bq}(Z')}{T(B_q)M_{Z'}^2}\left[
   C_1^{VLR}(\mu_{Z'})\langle Q_1^{VLR}(\mu_{Z'},B_q)\rangle+C_2^{VLR}(\mu_{Z'})\langle Q_2^{VLR}(\mu_{Z'},B_q)\rangle
 \right]~,\nonumber\\
\end{eqnarray}
where
\begin{eqnarray}
\gSM&\equiv&\frac{4G_F}{\sqrt{2}}\frac{\alpha}{2\pi\sin^2\theta_W}~,\\
 T(B_q)&\equiv&\frac{G_F^2}{12\pi^2}F_{B_q}^2{\hat B}_{B_q}m_{B_q}M_W^2(V_{tb}^*V_{tq})^2\eta_B~,
\end{eqnarray}
and ${\tilde r}=0.985$ for $M_{Z'}=1$ TeV.
For the scalar field, 
\begin{eqnarray}
  [\Delta S(B_q)]_{SLL}&=&
-\frac{[\Delta_L^{bq}(H)]^2}{T(B_q)2M_H^2}\left[
C_1^{SLL}(\mu_H)\langle Q_1^{SLL}(\mu_H, B_q)\rangle+C_2^{SLL}(\mu_H)\langle Q_2^{SLL}(\mu_H, B_q)\rangle
\right]~,\\
{[}\Delta S(B_q)]_{SRR}&=&[\Delta S(B_q)]_{SLL}(L\to R)~,\\
{[}\Delta S(B_q)]_{SLR}&=&
-\frac{\Delta_L^{bq}(H)\Delta_R^{bq}(H)}{T(B_q)M_H^2}\left[
C_1^{SLR}(\mu_H)\langle Q_1^{SLR}(\mu_H, B_q)\rangle+C_2^{SLR}(\mu_H)\langle Q_2^{SLR}(\mu_H, B_q)\rangle
\right]~.\nonumber\\
\end{eqnarray}
The expectation values of the operators $Q_i^a$ are
\begin{equation}
 \langle Q_i^a(\mu_M, B_q)\rangle\equiv\frac{1}{3}m_{B_q}F_{B_q}^2 P_i^a(\mu_M, B_q)~.
\end{equation}
For the case of $\Delta_R^{bq}=0$,
\begin{eqnarray}
 \frac{\Delta M_q(Z')}{\Delta M_q^{\rm SM}}
&=&\left| 1+\frac{[S(B_q)]_{VLL}}{S_0(x_t)}\right|\nn\\
&\simeq&
1+\frac{1}{S_0(x_t)}{\rm Re}\left[\frac{\Delta_L^{bq}(Z')}{V_{tb}^*V_{tq}}\right]^2
\frac{4{\tilde r}}{M_{Z'}^2\gSM^2}~,
\label{DMZ}
\end{eqnarray}
up to the leading order of $\Delta_L^{bq}$.
Now we define a double ratio $R_{\Delta M}^{Z'}$ as 
\begin{equation}
 R_{\Delta M}^{Z'}\equiv
\frac{\Delta M_s(Z')/\Delta M_s^{\rm SM}-1}{\Delta M_d(Z')/\Delta M_d^{\rm SM}-1}
=\frac{{\rm Re}\left[\Delta_L^{bs}(Z')/V_{ts}\right]^2}{{\rm Re}\left[\Delta_L^{bd}(Z')/V_{td}\right]^2}~,
\end{equation}
where the result of Eq.\ (\ref{DMZ}) is applied.
Similarly, for the scalar contribution (with $\Delta_R^{bq}=0$),
\begin{equation}
 R_{\Delta M}^H\equiv
\frac{\Delta M_s(H)/\Delta M_s^{\rm SM}-1}{\Delta M_d(H)/\Delta M_d^{\rm SM}-1}
=\frac{{\hat B}_{B_d}}{{\hat B}_{B_s}}
\frac{{\rm Re}\left[\Delta_L^{bs}(H)/V_{ts}\right]^2}{{\rm Re}\left[\Delta_L^{bd}(H)/V_{td}\right]^2}~.
\end{equation}
We assumed here that the light-quark dependence on $P_i^a(\mu_H,B_q)$ is negligible \cite{Urban},
and thus $P_i^a(\mu_H,B_d)\simeq P_i^a(\mu_H,B_s)$.
In the scalar unparticle scenario \cite{jplee1},
\begin{equation}
 \frac{\Delta M_q^\calU}{\Delta M_q^{\rm SM}}-1
\equiv|\Delta_\calU|-1
={\rm Re}\left[(c_{\calU L}^{bq})^2 f_\calU^q\cot\dU\pi\right]
+{\rm Im}\left[(c_{\calU L}^{bq})^2 f_\calU^q\right]+\calO(c_{\calU L}^4)~.
\label{DeltaU}
\end{equation}
Here
\begin{equation}
 f_\calU^q=
\frac{5}{24 M_{12}^{\rm SM}} A_\dU\left(\frac{F_{B_q}^2}{m_{B_q}}\right)
\left(\frac{m_{B_q}^2}{\Lambda_\calU}\right)^\dU~,
\end{equation}
where $M_{12}^{\rm SM}$ is the SM contribution and
\begin{equation}
 A_\dU\equiv
\frac{16\pi^{5/2}}{(2\pi)^{2\dU}}\frac{\Gamma(\dU+1/2)}{\Gamma(\dU-1)\Gamma(2\dU)}~,
\end{equation}
with $\dU$ being the scaling dimension of scalar unparticle operator.
The double ratio for scalar unparticle is 
\begin{eqnarray}
 R_{\Delta M}^\calU&\equiv&
\frac{\Delta M_s^\calU/\Delta M_s^{\rm SM}-1}{\Delta M_d^\calU/\Delta M_d^{\rm SM}-1}\nn\\
&\simeq&
\left(\frac{{\hat B}_{B_d}}{{\hat B}_{B_s}}\right)
\left(\frac{m_{B_s}^2}{m_{B_d}^2}\right)^{\dU-1}
\frac{{\rm Re}({\tilde c}_{\calU L}^{bs})^2\cot\dU\pi+{\rm Im}({\tilde c}_{\calU L}^{bs})^2}
{{\rm Re}({\tilde c}_{\calU L}^{bd})^2\cot\dU\pi+{\rm Im}({\tilde c}_{\calU L}^{bd})^2}~,
\end{eqnarray}
where we put $c_{\calU L}^{bq}\equiv{\tilde c}_{\calU L}^{bq}\cdot V_{tb}^*V_{tq}$.
For real ${\tilde c}_{\calU L}^{bq}$, one has
\begin{equation}
 R_{\Delta M}^\calU=
\left(\frac{{\hat B}_{B_d}}{{\hat B}_{B_s}}\right)
\left(\frac{m_{B_s}^2}{m_{B_d}^2}\right)^{\dU-1}
\left(\frac{{\tilde c}_{\cal U}^{bs}}{{\tilde c}_{\calU L}^{bd}}\right)^2~.
\label{RMU}
\end{equation}
If ${\tilde c}_{\calU L}^{bq}$ is purely imaginary, one gets a similar result.
\par
Now we move to $B_{d,s}\to\mu^+\mu^-$ decays.
The relevant effective Hamiltonian is given by
\begin{equation}
 \calH_{\rm eff}
=-\frac{G_F\alpha}{\sqrt{2}\pi}\left[
   V_{ts}^*V_{tb}\sum_i^{10,S,P}(C_i\calO_i+C_i'\calO_i')+{\rm h.c.}\right]~,
\label{Heff}
\end{equation}
where the operators $\calO_i$ are
\begin{eqnarray}
 \calO_{10}&=&(\sbar\gamma_\mu P_L b)({\bar\ell}\gamma^\mu\gamma_5\ell)~,~~~~~
 \calO_{10}'=(\sbar\gamma_\mu P_R b)({\bar\ell}\gamma^\mu\gamma_5\ell)~,\\
 \calO_S&=&m_b(\sbar P_R b)({\bar\ell}\ell)~,~~~~~~~~~~~
 \calO_S'=m_b(\sbar P_L b)({\bar\ell}\ell)~,\\
 \calO_P&=&m_b(\sbar P_R b)({\bar\ell}\gamma_5\ell)~,~~~~~~~~
 \calO_P'=m_b(\sbar P_L b)({\bar\ell}\gamma_5\ell)~.
\end{eqnarray}
\par
For $B_s$ decay it is convenient to define \cite{Bruyn}
\begin{equation}
 \Brbar(B_s\to\mu^+\mu^-)\equiv\frac{1}{r(y_s)}\Br(B_s\to\mu^+\mu^-)_{\rm th}~,
\end{equation}
where
\begin{eqnarray}
 r(y_s)&\equiv&\frac{1-y_s^2}{1+y_s\calA_{\Delta\Gamma}}~,\\
 y_s&\equiv&\tau_{B_s}\frac{\Delta\Gamma_s}{2}=0.088\pm0.014~,
\end{eqnarray}
and the asymmetric parameter 
\begin{equation}
 \calA_{\Delta\Gamma}\equiv\frac{R_H-R_L}{R_H+R_L}~,
\end{equation}
where $R_{H(L)}\exp\left[-\Gamma^{(s)}_{H(L)}t\right]$ is the decay rate of the heavy (light) mass eigenstate.
Here $\Br(B_s\to\mu^+\mu^-)_{\rm th}$ is a theoretical prediction 
while $\Brbar(B_s\to\mu^+\mu^-)$ would be directly compared with the experimental results.
In general,
\begin{equation}
 \frac{\Brbar(B_s\to\mu^+\mu^-)}{\Brbar(B_s\to\mu^+\mu^-)_{\rm SM}}
=\frac{1+y_s\calA_{\Delta\Gamma}}{1+y_s}\left(|P|^2+|S|^2\right)~,
\end{equation}
where
\begin{eqnarray}
 P&\equiv&
\frac{C_{10}-C_{10}'}{C_{10}^{\rm SM}}
+\frac{m_{B_s}^2}{2m_\mu}\frac{m_b}{m_b+m_s}\frac{C_P-C_P'}{C_{10}^{\rm SM}}
\equiv |P|e^{i\varphi_P}~,\\
S&\equiv&
\sqrt{1-\frac{4m_\mu^2}{m_{B_s}^2}}~\frac{m_{B_s}^2}{2m_\mu}\frac{m_b}{m_b+m_s}\frac{C_S-C_S'}{C_{10}^{\rm SM}}
\equiv |S|e^{i\varphi_S}~.
\end{eqnarray}
The standard model contribution is
\begin{equation}
 C_{10}^{\rm SM}=-\frac{1}{\sin^2\theta_W}\eta_Y Y_0(x_t)~,
\end{equation}
with $\eta_Y=1.012$ and 
\begin{equation}
 Y_0(x_t)
=\frac{x_t}{8}\left[
  \frac{x_t-4}{x_t-1}+\frac{3x_t\log x_t}{(x_t-1)^2}\right]~.
\end{equation}
\par
For $Z'$ model,
\begin{eqnarray}
 \sin^2\theta_W C_{10}(Z')&=&
-\eta_Y Y_0(x_t)-\frac{1}{\gSM^2}\frac{1}{M_{Z'}^2}
  \frac{\Delta_L^{sb}(Z')\Delta_A^{\mu\mu}(Z')}{V_{ts}^*V_{tb}}~,\\
 \sin^2\theta_W C_{10}'(Z')&=&
-\frac{1}{\gSM^2}\frac{1}{M_{Z'}^2}
  \frac{\Delta_R^{sb}(Z')\Delta_A^{\mu\mu}(Z')}{V_{ts}^*V_{tb}}~,
\end{eqnarray}
while other coefficients are vanishing.
Using $\Delta_{L,R}^{sb}(Z')=\Delta_{L,R}^{bs}(Z')^*$, one has
\begin{eqnarray}
 \frac{\Brbar(B_s\to\mu^+\mu^-)}{\Brbar(B_s\to\mu^+\mu^-)_{\rm SM}}-1&\simeq&
\frac{y_s}{1+y_s}\left[\cos(2\theta_Y^{B_s}+\theta_S^{B_s})-1\right]\nn\\
&&
+\frac{1}{1+y_s}\frac{1}{\eta_Y Y_0(x_t)}\frac{1}{M_{Z'}^2\gSM^2}
  2{\rm Re}\left[\frac{(\Delta_L^{bs*}-\Delta_R^{bs*})\Delta_A^{\mu\mu}}{V_{ts}^*V_{tb}}\right]~,
\end{eqnarray}
and
\begin{equation}
 \frac{\Brbar(B_d\to\mu^+\mu^-)}{\Brbar(B_d\to\mu^+\mu^-)_{\rm SM}}-1\simeq
\frac{1}{\eta_Y Y_0(x_t)}\frac{1}{M_{Z'}^2\gSM^2}
  2{\rm Re}\left[\frac{(\Delta_L^{bd*}-\Delta_R^{bd*})\Delta_A^{\mu\mu}}{V_{td}^*V_{tb}}\right]~,
\end{equation}
up to $\calO(y_s\Delta_{L,R}\Delta_A)$.
For $\Delta_R=0$ and $\Delta_L^{bq}={\tilde\Delta}_L^{bq} V_{tq}$ where ${\tilde\Delta}_L^{bq}$ is real, 
the double ratio 
\begin{equation}
 R_{\mu\mu}^{Z'}\equiv
\left[
\frac{\Brbar(B_s\to\mu^+\mu^-)_{Z'}}{\Brbar(B_s\to\mu^+\mu^-)_{\rm SM}}-1\right]/
\left[
\frac{\Br(B_d\to\mu^+\mu^-)_{Z'}}{\Br(B_d\to\mu^+\mu^-)_{\rm SM}}-1\right]
\label{RmumuZ}
\end{equation}
remarkably reduces to
\begin{equation}
 R_{\mu\mu}^{Z'}=\frac{1}{1+y_s}\left(\frac{{\tilde\Delta}_L^{bs}}{{\tilde\Delta}_L^{bd}}\right)~.
\end{equation}
In this case the ratio $R_{\Delta M}^{Z'}=({\tilde\Delta}_L^{bs}/{\tilde\Delta}_L^{bd})^2$, 
and thus one arrives at a very simple relation
\begin{equation}
 R_{\mu\mu}^{Z'}(1+y_s)=\sqrt{R_{\Delta M}^{Z'}}~.
\end{equation}
\par
For neutral scalar $H$, the coefficients are
\begin{eqnarray}
 C_{10}(H)&=&C_{10}^{\rm SM}~,\\
 C_S(H)&=&\frac{1}{m_b\sin^2\theta_W}\frac{1}{\gSM^2}\frac{1}{M_H^2}
  \frac{\Delta_R^{sb}(H)\Delta_S^{\mu\mu}(H)}{V_{ts}^*V_{tb}}~,\\
 C_S'(H)&=&\frac{1}{m_b\sin^2\theta_W}\frac{1}{\gSM^2}\frac{1}{M_H^2}
  \frac{\Delta_L^{sb}(H)\Delta_S^{\mu\mu}(H)}{V_{ts}^*V_{tb}}~,\\
 C_P(H)&=&\frac{1}{m_b\sin^2\theta_W}\frac{1}{\gSM^2}\frac{1}{M_H^2}
  \frac{\Delta_R^{sb}(H)\Delta_P^{\mu\mu}(H)}{V_{ts}^*V_{tb}}~,\\
 C_P'(H)&=&\frac{1}{m_b\sin^2\theta_W}\frac{1}{\gSM^2}\frac{1}{M_H^2}
  \frac{\Delta_L^{sb}(H)\Delta_P^{\mu\mu}(H)}{V_{ts}^*V_{tb}}~.
\end{eqnarray}
One can define a double ratio $R_{\mu\mu}^H$ similar to Eq.\ (\ref{RmumuZ}).
For simplicity we assume that $\Delta_R=0$ and 
$\Delta_L^{bq}={\tilde\Delta}_L^{bq} V_{tq}$ with real ${\tilde\Delta}_L^{bq}$.
Note that in this case 
\begin{equation}
 R_{\Delta M}^H=\frac{{\hat B}_{B_d}}{{\hat B}_{B_s}}
\left(\frac{{\tilde\Delta}_L^{bs}}{{\tilde\Delta}_L^{bd}}\right)^2~.
\end{equation}
For the case of $\Delta_S^{\mu\mu}(H)=0$, the double ratio reduces to be
\begin{eqnarray}
 R_{\mu\mu}^H&\equiv&
\left[
\frac{\Brbar(B_s\to\mu^+\mu^-)_H}{\Brbar(B_s\to\mu^+\mu^-)_{\rm SM}}-1\right]/
\left[
\frac{\Br(B_d\to\mu^+\mu^-)_H}{\Br(B_d\to\mu^+\mu^-)_{\rm SM}}-1\right]\nn\\
&=&
\frac{1}{1+y_s}\left(
\frac{m_{B_s}^2}{m_{B_d}^2}\frac{m_b+m_d}{m_b+m_s}\right)^2
\left(\frac{{\hat B}_{B_s}}{{\hat B}_{B_d}}\right)
R_{\Delta M}^H~.
\label{RmumuH}
\end{eqnarray}
On the other hand if $\Delta_P^{\mu\mu}=0$, 
\begin{equation}
 R_{\mu\mu}^H=
\left(\frac{1-2y_s}{1+y_s}\right)
\left(\frac{1-4m_\mu^2/m_{B_s}^2}{1-4m_\mu^2/m_{B_d}^2}\right)
\left(\frac{m_{B_s}^2}{m_{B_d}^2}\frac{m_b+m_d}{m_b+m_s}\right)^2
\left(\frac{{\hat B}_{B_s}}{{\hat B}_{B_d}}\right)
R_{\Delta M}^H
\label{RmumuH2}
\end{equation}
\par
For scalar unparticles \cite{jplee2}, 
\begin{eqnarray}
 P&=&1-\frac{\sin^2\theta_W}{\eta_Y Y_0(x_t)}\frac{\sqrt{2}\pi A_\dU}{\alpha G_F m_{B_s}^2}
\left(\frac{m_{B_s}}{\Lambda_\calU}\right)^{2\dU}
\left(\frac{m_b}{m_b+m_s}\right)
\left(\frac{c_{\calU L}^{bs}c_{\calU L}^\ell}{V_{tb}^*V_{ts}}\right)^*
(\cot\dU\pi+i)~,\\
S&=&0~,
\end{eqnarray}
and thus $\calA_{\Delta\Gamma}=\cos(2\varphi_P-\phi_s^\calU)$.
Here $\phi_s^\calU$ is the phase of $\Delta_\calU$ in Eq.\ (\ref{DeltaU}).
For real ${\tilde c}_{\calU L}^{bq}, c_{\calU L}^\ell$, 
$\cos(2\varphi_P-\phi_s^\calU)\simeq 1$ up to $\calO(c_{\calU L})^4$,
and the double ratio is
\begin{eqnarray}
 R_{\mu\mu}^\calU&\equiv&
\left[
\frac{\Brbar(B_s\to\mu^+\mu^-)_\calU}{\Brbar(B_s\to\mu^+\mu^-)_{\rm SM}}-1\right]/
\left[
\frac{\Br(B_d\to\mu^+\mu^-)_\calU}{\Br(B_d\to\mu^+\mu^-)_{\rm SM}}-1\right]\nn\\
&=&
\frac{1}{1+y_s}\left(\frac{m_{B_s}}{m_{B_d}}\right)^{2\dU-2}
\left(\frac{m_b+m_d}{m_b+m_s}\right)
\left(\frac{{\tilde c}_{\calU L}^{bs}}{{\tilde c}_{\calU L}^{bd}}\right)\nn\\
&=&
\frac{1}{1+y_s}\left(\frac{m_{B_s}}{m_{B_d}}\right)^{\dU-1}
\left(\frac{m_b+m_d}{m_b+m_s}\right)
\sqrt{\frac{{\hat B}_{B_s}}{{\hat B}_{B_d}}}
\sqrt{R_{\Delta M}^\calU}~,
\end{eqnarray}
where the result of Eq.\ (\ref{RMU}) is used.
Our results are summarized as follows:
\begin{eqnarray}
 R_{\mu\mu}^{Z'}(1+y_s)&=&\sqrt{R_{\Delta M}^{Z'}}~,
\label{result1}\\
  R_{\mu\mu}^H(1+y_s)&=&
\left(
\frac{m_{B_s}^2}{m_{B_d}^2}\frac{m_b+m_d}{m_b+m_s}\right)^2
\left(\frac{{\hat B}_{B_s}}{{\hat B}_{B_d}}\right)
R_{\Delta M}^H~({\rm if}~\Delta_S^{\mu\mu}=0)~,
\label{result2}\\
 R_{\mu\mu}^H(1+y_s)&=&
\left(1-2y_s\right)
\left(\frac{1-\frac{4m_\mu^2}{m_{B_s}^2}}{1-\frac{4m_\mu^2}{m_{B_d}^2}}\right)
\left(\frac{m_{B_s}^2}{m_{B_d}^2}\frac{m_b+m_d}{m_b+m_s}\right)^2
\left(\frac{{\hat B}_{B_s}}{{\hat B}_{B_d}}\right)
R_{\Delta M}^H~({\rm if}~\Delta_P^{\mu\mu}=0)~,
\label{result3}\\
 R_{\mu\mu}^\calU(1+y_s)&=&
\left(\frac{m_{B_s}}{m_{B_d}}\right)^{\dU-1}
\left(\frac{m_b+m_d}{m_b+m_s}\right)
\sqrt{\frac{{\hat B}_{B_s}}{{\hat B}_{B_d}}}
\sqrt{R_{\Delta M}^\calU}~.
\label{result4}
\end{eqnarray}
The reason why $R_{\mu\mu}^H\sim R_{\Delta M}^H$ is that in $R_{\mu\mu}^H$, 
$\Br/\Br_{\rm SM}-1$ is non-vanishing only at $\calO(c^2)$, due to the fact that $\Delta_P^{\mu\mu}$ is pure imaginary \cite{Buras3}.
\par
Numerically, Eqs.\ (\ref{result1})-(\ref{result4}) are 
\begin{eqnarray}
\label{numerics}
 R_{\mu\mu}^{Z'}&=&0.919\times\sqrt{R_{\Delta M}^{Z'}}=0.775~,\\
 R_{\mu\mu}^{H,\Delta_S=0}&=&0.993\times R_{\Delta M}^H=0.707~,\\
 R_{\mu\mu}^{H,\Delta_P=0}&=&0.818\times R_{\Delta M}^H=0.583~,\\
 R_{\mu\mu}^\calU&=&(1.02)^{\dU-1}\times0.925\times\sqrt{R_{\Delta M}^\calU}=0.780\times(1.02)^{\dU-1}~,
\end{eqnarray}
where $R_{\Delta M}=0.712$ is used. 
The above results can be used to predict the yet-to-be-measured branching ratio, $\Br(B_d\to\mu^+\mu^-)$.
Table \ref{table1} shows the predicted values of $\Br(B_d\to\mu^+\mu^-)$.
\begin{table}
 \begin{tabular}{c||c|c|c|c}
New Physics & $Z'$ & $H$($\Delta_S^{\mu\mu}=0$) & $H$($\Delta_P^{\mu\mu}=0$) & $\calU(\dU=1.5)$\\\hline
$\Brbar(B_s\to\mu^+\mu^-)=2.9\times10^{-10}$ \cite{LHCb1307}
& $0.799\times 10^{-10}$ & $0.775\times 10^{-10}$ & $0.716\times 10^{-10}$ & $0.803\times 10^{-10}$\\
$\Brbar(B_s\to\mu^+\mu^-)=3.0\times10^{-10}$ \cite{CMS1307} 
& $0.837\times 10^{-10}$ & $0.816\times 10^{-10}$ & $0.766\times 10^{-10}$ & $0.840\times 10^{-10}$\\
$\Brbar(B_s\to\mu^+\mu^-)=3.2\times10^{-10}$ \cite{LHCb1211}
& $0.913\times 10^{-10}$ & $0.900\times 10^{-10}$ & $0.868\times 10^{-10}$ & $0.915\times 10^{-10}$ 
 \end{tabular}
\caption{Predictions for $\Br(B_d\to\mu^+\mu^-)$ for various $\Br(B_s\to\mu^+\mu^-)$ measurements.
For unparticles, the branching ratio is given at a reference point $\dU=1.5$.}
\label{table1}
\end{table}
Note that the values of Table \ref{table1} are all far below the current upper bound, 
$\Br(B_d\to\mu^+\mu^-)<7.4\times 10^{-10}$ by the LHCb \cite{LHCb1307} 
and $\Br(B_d\to\mu^+\mu^-)<1.1\times 10^{-9}$ by the CMS \cite{CMS1307},
and slightly smaller than the SM prediction, $\Br(B_d\to\mu^+\mu^-)_{\rm SM}=1.05\times 10^{-10}$.
This is because $\Brbar(B_s\to\mu^+\mu^-)<\Brbar(B_s\to\mu^+\mu^-)_{\rm SM}=(3.56\pm0.18)\times 10^{-9}$ \cite{Buras3}
and $R_{\Delta M}=0.712>0$.
Note also that the predictions are made {\em without knowing} any numerical details of the new couplings
except that they are small enough to neglect higher orders.
In this way, by measuring $\Br(B_d\to\mu^+\mu^-)$ we can easily figure out which kind of new physics is realized in $B$ systems.
\par
In conclusion, we derived new relations between $B_{d,s}$ observables.
The relations are valid only when new physics exists in $B_{d,s}$ systems, 
which is a very plausible assumption.
The relations are different in specific models.
In this analysis we only consider flavor changing scalar (un)particles and vector bosons.
For other models one can define similar double ratios as given in this work.
The double ratios become very simple when there are only left-(or right-)handed couplings, 
and the couplings are MFV-like.
If this were not the case, then our simple relations would not hold any more.
In other words, if we confirm that the simplified double ratio relations really hold, 
then we may conclude that the new physics is realized in a minimal way.
\par 
One point to be mentioned is that our double ratio becomes meaningless if there were no new physics at all.
In this case both numerator and denominator are vanishing and one cannot take a ratio.
Thus the double ratio is not adequate to check whether there is any new physics or not,
but to see which kind of NP is involved once the observables are turned out to be quite different from the SM predictions.
Current status of NP searches in $B$ meson is not so pessimistic.
According to \cite{Charles}, the relative size of NP in $\Delta M_{d,s}$ ($=h_{d,s}$) is currently $\lesssim 0.2 \sim 0.3$,
and would be $\lesssim 0.1$ in near future ("Stage I" where the LHCb will end).
As for $B_d\to\mu^+\mu^-$, current upper bound is almost order of magnitude larger than the SM prediction.
It is predicted in \cite{Altmannshofer} that at $2\sigma$,
$0.3\times10^{-10}\lesssim{\rm Br}(B_d\to\mu^+\mu^-)\lesssim1.8\times 10^{-10}$.
If the measured branching ratio does not lie within this window, it would be a clear indication of NP.
It is also found in \cite{Altmannshofer} that although the measured value of $\Br(B_s\to\mu^+\mu^-)$ provide constraints on NP,
there are still sizable regions allowed for $C_S-C_S'$ and $C_P-C_P'$ parameter space.
\par
Besides the current status of NP searches, we need NP for various reasons (dark matter for example).
Although there have been no smoking-gun signals for NP up to now, 
we believe that the SM is not (and should not be) the full story of particle physics.
In this context the double ratio analysis might be very promising with the coming flavor precision era,
and can be also applied to the $K$ meson systems.
\begin{acknowledgments}
This work is supported by WCU program through the KOSEF funded by the MEST (R31-2008-000-10057-0).
\end{acknowledgments}

\end{document}